\documentclass[doublecol]{epl2} 

\usepackage{graphicx, xcolor, amsfonts}

\def\calc{\mathcal{C}}
\def\calM{\mathcal{M}}
\def\calm{\mathfrak{m}}
\def\pin{P_{\textrm{\scriptsize{in}}}}
\def\pout{P_{\textrm{\scriptsize{out}}}}
\def\pinout{P_{\textrm{\scriptsize{in,out}}}}

\title{Harmonic structures of Beethoven quartets: a complex network approach}

\author{T. Frottier\inst{1} \and B. Georgeot\inst{1} \and O. Giraud\inst{2}}
\shortauthor{T. Frottier\etal}

\institute{                    
  \inst{1}Laboratoire de Physique Th\'eorique, Universit\'e de Toulouse, CNRS, UPS, 31062 Toulouse, France\\
\inst{2} Universit\'e Paris Saclay, CNRS, LPTMS, 91405, Orsay, France
}

\abstract{We propose a complex network approach to the harmonic structure underpinning western tonal music. From a database of Beethoven's string quartets, we construct a directed network whose nodes are musical chords and edges connect chords following each other. We show that the network is scale-free and has specific properties when ranking algorithms are applied. We explore its community structure and its musical interpretation, and propose statistical measures stemming from network theory allowing to distinguish stylistically between periods of composition. Our work opens the way to a network approach of structural properties of tonal harmony.}

\begin{document}

\maketitle

\section{Introduction}

In the recent past, network theory has been developed as a new tool enabling to uncover structural properties, dynamics and evolution of a variety of systems, from natural ones, such as biological networks, to human produced ones such as the World Wide Web or social networks \cite{boccaletti2006complex}. Interestingly enough, it has also been shown that this theory can give new insights on systems which are not obviously organized as networks, such as languages \cite{cancho2001small, masucci2006network, antiqueira2007strong, sole2010language} or board games \cite{georgeot2012game, kandiah2014move, xu2015weiqi, coquide2017distinguishing}. 

Music shares features from both languages and games. Connections between music and natural sciences are numerous, be it physiology, physics of waves, or group theory \cite{benson2006music}.
The improvement of computer capabilities has recently opened several lines of research at the interface between musicology, mathematics and computer science, from automatic harmonic analysis \cite{de2013automatic, kroger2008rameau}, statistical analysis of music \cite{rohrmeier2008statistical}, creation of databases \cite{neuwirth2018annotated, hentschel2021annotated}, computer-assisted Schenkerian analysis \cite{marsden2010schenkerian}, to recent applications from machine learning \cite{elgammal, landsnes2019model}. A huge corpus of musical pieces exists with many musicological studies analyzing their history and evolution
(see e.g.~\cite{beach1974origins,darrigol2007acoustic} for a historical account). 
Importantly, musical syntax is not so much about the perception of isolated chords as about the relationship between a chord and the ones that surround it. An important aspect of musical analysis is thus to understand how chords are interrelated, both at a global and at a local scale, the latter corresponding to the neighbourhood of a chord in a given musical segment. In this paper, our aim is to apply network theory to musical pieces. 

In 2018, a database of all chords of Beethoven string quartets was established \cite{neuwirth2018annotated}, based on harmonic analyses made by human experts from the musical scores of the quartets. This database was analysed in \cite{moss2019statistical}, where the authors investigated the frequency distribution of chords, pairs of chords, and higher-order $n$-grams.

In the present paper we go a step further by constructing a network based on temporal relations between chords within musical segments, following ideas borrowed from text analysis \cite{cancho2001small, masucci2006network, antiqueira2007strong, sole2010language}. In the following, we build the network, discuss its properties and their relationship with musical features, and investigate its community structure. The string quartets are particularly interesting in that their composition stretches over a period of 28 years of Beethoven's life, allowing to follow the stylistic evolution throughout his lifetime \cite{ratner1970key, knittel1998wagner, bonds2017irony}. 
Here we show that tools from network theory allow to statistically differentiate between the different periods of the Beethoven quartets.

The present network approach bears some analogy with the Euler Tonnetz, the geometry of musical chords \cite{tymoczko2006geometry}, or a network approach of atonal music in \cite{buongiorno2020hitchhiker}. A network of chord progressions, similar in spirit to ours, was proposed very recently in \cite{buongiorno2021tonal}, based on chords taken as vertical arrangements of pitch classes and a small-scale analysis of data. By contrast, our approach considers chords in a functional relationship with a local key, as determined by human experts, and over the scale of a whole corpus. Our work shows that a network approach provides some insight into structural properties of tonal harmony.

\section{The database}

The annotated database of all chords from the complete set of Beethoven string quartets (16 quartets, a total of 70 movements) is available online at \cite{abc}. It has 28095 entries, each of which provides information on a chord : the global and local key with respect to which it appears, as well as possible changes in the chord, relative root, or pedals. It also contains information on the chord duration, movement and measure in which it appears, and whether or not it is at the end of a musical segment.

Each chord is characterized by a Latin numeral (from I to VII) indicating its relation with the local key, and a figure (6, 64, 7, 65, 43, 2) indicating whether the chord is in root position or appears as an inversion. It may also contain an indication of its form (major seventh, half-diminished, diminished and augmented, respectively denoted M, $\%$, $\circ$, and $+$), various figures between brackets indicating changes in the inversion, and/or additional Latin numbers indicating the relative root. For some of the chords (107 of them in the database), no harmonic value could be determined by the experts, and a label "none" was assigned; such chords are most frequent in the late quartets (with 58 undetermined chords), which are known to be more innovative.

The whole corpus is divided into 929 musical segments. Each has a specific local key, which indicates whether the segment is major or minor. We split the database into two parts, one for chords in segments with major local key and one for chords in minor local key. Among the 929 musical segments there are 551 major and 378 minor ones, yielding a database of 20276 chords in major segments and 7819 chords in minor segments. Among these 20276 entries in major segments we found $N_{\calM}=871$ distinct ones. The minor segments involve $N_m=599$ distinct entries. 

Individual chords can be ranked by their number of occurrences in the database. We found that the frequency $f$ of occurences as a function of the rank $r$ has a power-law tail $f\propto 1/r^{\gamma}$. This is characteristic of the Zipf law, which was first observed in the analysis of languages \cite{zipf1935psycho} and since then in many contexts. For our database, a similar power-law tail was already obtained in \cite{moss2019statistical}. Here we find an exponent $\gamma\approx 1.6$, not so far from the exponent $\approx 2$ found in natural languages.

\section{Network theory}

To go beyond mere statistics of chords, we now introduce the basic tools of network theory.

\subsection{The networks}
A graph is a set of vertices connected by edges. In our case, we construct a graph $\calM$ based on chords from major segments. Its vertices are the $N_{\calM}$ distinct chords appearing in the database. We also construct a graph $\calm$  with chords from minor segments, yielding a graph with $N_\calm$ vertices.

\begin{figure}
\begin{center}
\includegraphics[width=1.\linewidth]{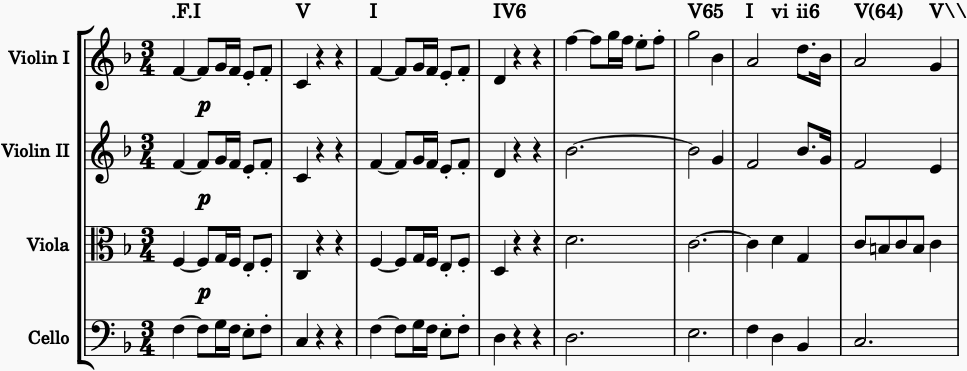}\\
\vspace{.4cm}
\includegraphics[width=0.59\linewidth]{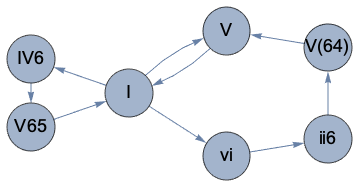}
\end{center}
\caption{First segment of Beethoven's Op.\,18 No.\,1 in F major (first movement) and its associated graph. The score was generated from the data of \cite{abc} using MuseScore \cite{musescore}.}
\label{exemplescore}
\end{figure}

Our networks are built in the following way: We add a directed edge between two vertices $i$ and $j$ each time chord $j$ immediately follows chord $i$ in  the same segment. There are as many edges between $i$ and $j$ as occurrences of the pair $i,j$ in the database, which makes our graph a weighted directed graph. As an illustration, in Fig.~\ref{exemplescore} we show the first 8 bars of Beethoven's Op.\,18 No.\,1, which correspond to the first segment of the database. It consists of 10 labeled chords (7 distinct ones). The corresponding graph, with 7 vertices and 9 edges, is given below the score.

\subsection{The PageRank algorithm}
One of the tools developed for investigating the network structure is the PageRank algorithm, which gave the original impulse to the development of the Google search engine \cite{brin1998anatomy}. The PageRank algorithm is built to hierarchize the nodes of a network in a relevant way, by constructing a vector (the PageRank vector) whose entries are used to rank the vertices by order of importance. This vector is the eigenvector associated with the largest eigenvalue of a matrix $G$ constructed from the $N\times N$ adjacency matrix of the network. 

This Google matrix $G$ is defined, for some parameter $\alpha$ in $[0,1]$, as $G_{ij}=\alpha S_{ij}+(1-\alpha)/N$, where $S$ is obtained from the weighted adjacency matrix by replacing any column containing only 0 by a column of $1/N$ and normalizing the sum of entries of each column to 1. In the case of the quartet database, we did not encounter any such column of zeros, as segments end up with chords which are frequent in the database. Adding the constant part proportional to $1-\alpha$ to that matrix $S$ avoids numerical results being dominated by dangling groups (that is, groups of vertices with no outgoing edges), as those tend to dominate the PageRank when $\alpha \rightarrow 1$. 

By construction, $G$ is a stochastic matrix. Perron-Frobenius theorem ensures that $G$ has an eigenvector with eigenvalue 1 and real positive entries. The PageRank vector $p$ is defined as the vector such that $Gp=p$ and $\sum_i p_i=1$. The value $p_i$ can be interpreted as the probability for a random surfer following the edges of the network for an infinite time to be found on vertex $i$, if at each step the outgoing edges are chosen at random with equal probability. Properties of the Google matrix shed light on the structure of the network; this approach was successfully applied in a number of contexts \cite{ermann2015google}.

\subsection{Network communities}
The most basic structure that underlies the topology of a graph is its partition into communities, that is, subsets that have more connections within themselves than between one another. A way of determining whether a given partition of the set of vertices properly describes its community structure is to compute the modularity of that partition. This quantity measures how far a given graph is from a graph with the same connectivity but with edges taken at random within and between subsets of the partition. For a given partition into communities $\{\calc\}$, the modularity is given by 
$\mu=\sum_\calc\sum_{i,j\in\calc}[a_{ij}-d_id_j/(2m)]$, where $a_{ij}$ is equal to the number of undirected edges connecting vertices $i$ and $j$, $d_i$ is the total number of edges from $i$, $m$ is the total number of edges in the undirected graph, and the sum runs over all communities $\calc$. 
The partition that yields the highest modularity provides a possible decomposition of the graph into communities \cite{newman2004finding, fortunato2010community}.

\section{The musical network}
\begin{figure}[!t]
\begin{center}
\includegraphics[width=1\linewidth]{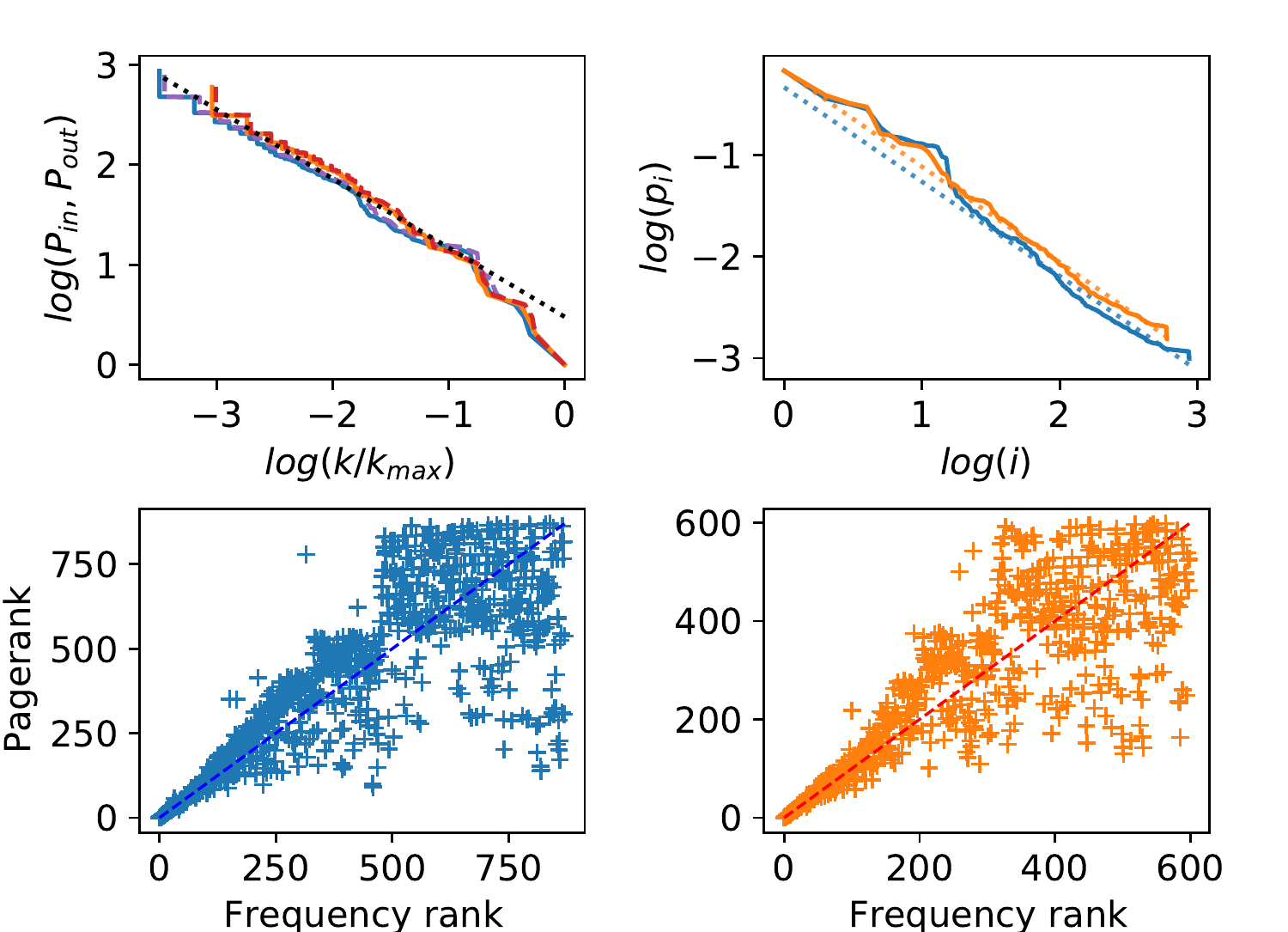}\hfill
\end{center}
\caption{Top left: Cumulative distribution of ingoing edges $\pin$ (major blue, minor orange) and outgoing edges $\pout$ (major violet, minor red, almost indistinguishable from $\pin$). A linear fit over the 30 leftmost points yields slopes -0.68 for major $\pinout$ and -0.70 for minor $\pinout$. Dotted black line has a slope -0.69. Top right: PageRank vector $p_i$ (sorted in descending order) for graphs $\calM$ (blue) and $\calm$ (orange). Dotted lines are linear fit, with slope -0.93 (major) and -0.94 (minor). Bottom: PageRank vs frequency rank for the major (left, blue) and minor (right, orange). Each point represents a specific chord. Dashed line is the line $y=x$.}\label{fig2}
\end{figure} 

We now apply the above tools to our musical networks. In Fig.~\ref{fig2} (top left panel) we display the cumulative distribution of incoming and outgoing edges, that is, the number of vertices that have more than $k$ ingoing (or outgoing) edges, with $k$ normalized by its maximum value. It follows a power-law $\pinout\sim 1/k^\nu$ with exponent $\nu\approx 0.7$. Similar power-law distributions of vertex connectivities were found in many real-world complex networks, known as scale-free networks \cite{barabasi1999emergence, albert2002statistical}. Here the exponent $1+\nu$ of the (non-integrated) distribution roughly corresponds to the exponent $\gamma=1.6$ found for the Zipf law. Typically in scale-free networks the exponent $1+\nu$ ranges from 2 to 3, but lower exponents have been found, for example $\approx 1.5$ for e-mail networks \cite{ebel2002scale}.

\subsection{PageRank vector}
In the top right panel of Fig.~\ref{fig2} we show the ranking of chords as given by the PageRank vector. As shown in the lower panels of Fig.~\ref{fig2}, this ranking is quite different from that given by the mere frequency of chords in the database. Some chords, although rare in the database, have a high PageRank; they can correspond to rare followers of much more common patterns. This is the case, for instance, for the chord labeled bIII, which generally follows the high-rank chord I.
The PageRank vector follows a power-law $p_i\sim 1/i^\beta$ with $\beta\approx 0.93$, very close to the exponent $0.9$ found for networks describing parts of the World Wide Web \cite{donato2004large, pandurangan2006using, giraud2009delocalization, georgeot2010spectral}.

\subsection{Spectrum of the Google matrix}
\begin{figure}
\begin{center}
\includegraphics[width=0.66\linewidth]{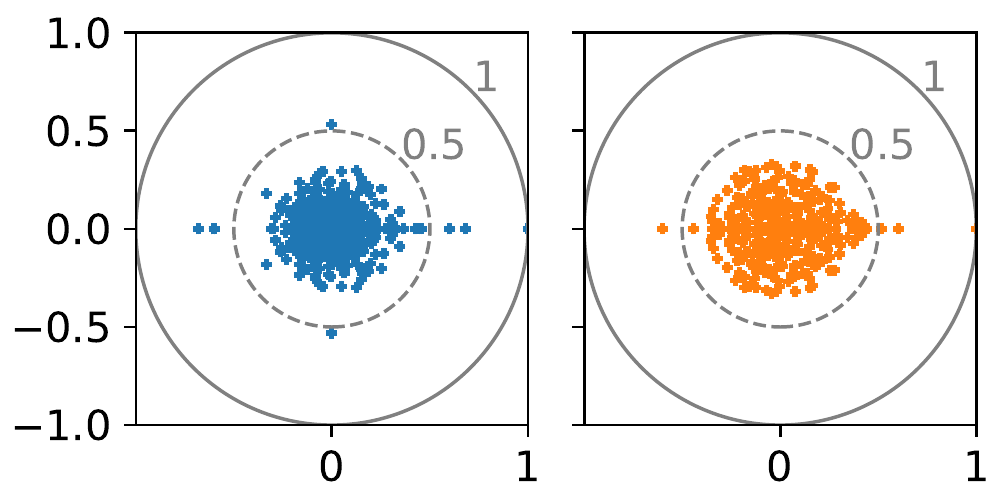}\hfill
\end{center}
\caption{Spectrum of $G$ in the complex plane (left major, right minor) for $\alpha=0.85$. Dashed circle of radius $\frac12$ is an eyeguide.}
\label{spectre}
\end{figure}

The spectrum of $G$ gives some insight into the structure of the network. 
For a symmetric matrix the spectrum is real. For directed networks
the matrix $G$ is in general non-symmetric, and the complex spectrum is all the more flattened onto the real axis as there exist pairs of edges of opposite directions between pairs of vertices. For example, this happens for dictionary networks, where many words are symmetrically related \cite{georgeot2010spectral}. The spectrum of $G$, displayed in Fig.~\ref{spectre}, shows that there is no such phenomenon, consistently with the temporal directionality of music. For instance, in major segments there are in total 40 occurences of the pair ii$\to$V, but only 7 of the pair V$\to$ii. 
This aspect of music, referred to as directedness, is also discussed in \cite{moss2019statistical}.

From Perron-Frobenius theorem, the spectrum of $G$ is by construction bounded by the circle of radius $\alpha$, except for the lone eigenvalue corresponding to the PageRank. However, it is clear from Fig.~\ref{spectre} that the spectrum is concentrated much closer to the center than the theoretical bound, almost entirely within a circle of radius $\frac12$. This reflects the fact that the network is highly connected, with many edges between different parts. Indeed, eigenvalues with large modulus correspond to long-lived eigenstates located on parts of the network which are less connected with the rest. For a dense graph, isolated regions are rarer, which tends to suppress such outlying eigenvalues.
In the present case, the smaller radius of the spectrum can be interpreted as a reflection of the fact that the same chords can appear in many different contexts, which homogeneizes the graph. The spectrum of $\calm$ is less concentrated, indicating that this phenomenon is less pronounced in minor. 
Lastly, we can notice in Fig.~\ref{spectre} the presence of isolated eigenvalues separated from the main cluster of eigenvalues. As will be illustrated below on the spectra of Fig.~\ref{spectreparperiode}, this reflects zones of the graph where groups of nodes are weakly coupled to other parts (for example, patterns of chords that appear only in specific contexts). Such features correspond to the notion of communities in a graph, to which we now turn.

\subsection{Communities}

\begin{figure}[!t]
\begin{center}
  \includegraphics[width=1\linewidth]{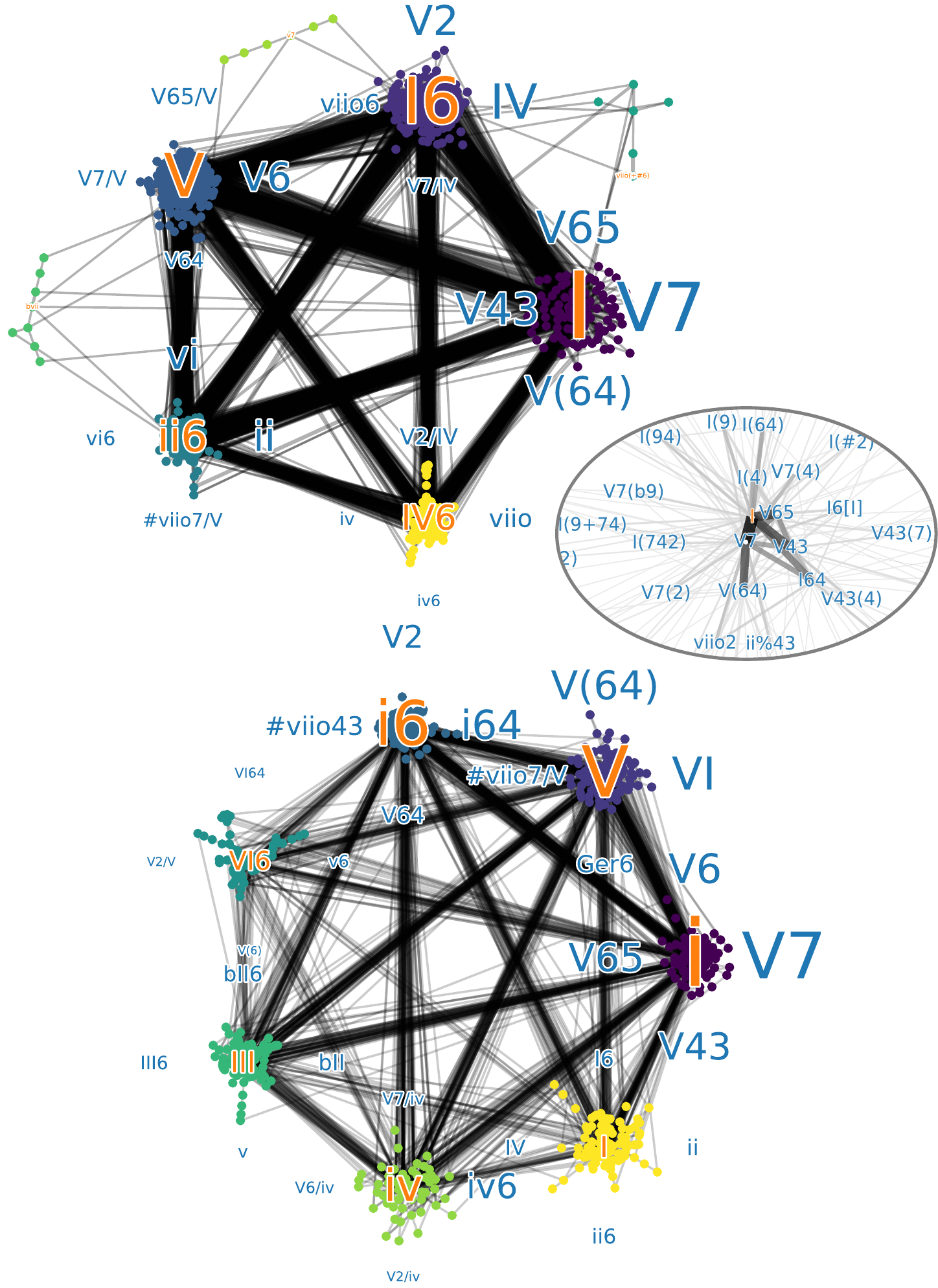}
  \end{center}
\caption{Community partition of the networks (top major, bottom minor). At the centre of each cloud (in orange) is the chord with largest $p_i$ (highest PageRank) in the community; the next four chords in PageRank order are labeled in blue around each community. Symbol size reflects value of $p_i$. Inset: zoom on the community I of $\calM$.}
\label{figcommunautes}
\end{figure}

As mentioned earlier, communities can be obtained by computing the modularity of partitions of the undirected graph and identifying the partition of maximal modularity. There is a variety of ways of computing the partition with highest modularity. We use the Louvain algorithm \cite{blondel2008fast}, implemented in the NetworkX package of Python. The output of the algorithm depends on a random seed, and for a given graph the resulting maximal modularity changes (mildly) from one run of the program to the other, as well as the partition itself. Nevertheless, the main features of the communities are robust. 

In Fig.~\ref{figcommunautes} we show the community partition for our graphs. We illustrate our results with the outcome with the highest modularity, namely 0.2252 for the graph $\calM$ and 0.2572 for the graph $\calm$. 
For the graph $\calM$, we find 5 main communities, each of which revolves around an elementary chord: I, IV, V, I6 and ii6. Within these communities one finds closely related chords. For instance, the community "IV" localizes on the subdominant IV but also on chords which have a close harmonic function with respect to the subdominant (such as V2/IV, V7/IV). The small outlying communities, very weakly connected with the rest of the graph, correspond to sequences of rare chords (such as chords appearing only once in the corpus). 

This community structure reflects the presence of poles of attraction, a dimension of music referred to as centricity in \cite{moss2019statistical}. As can be expected from a musical perspective, these poles include (in major) the fourth degree IV, dominant V, and tonic I. But interestingly, inversions of V belong to the same community as I. Moreover the other poles of attraction are also surrounded by inversions of their relative dominant, which  shows that they locally behave as the tonic. Similar features are found in the graph $\calm$ in Fig.~\ref{figcommunautes} bottom. The partition into communities thus yields a mesoscopic picture and allows to assess the role of chords within a community.

\section{Comparison between the different periods}

 \begin{figure}
\begin{center}
\includegraphics[width=1\linewidth]{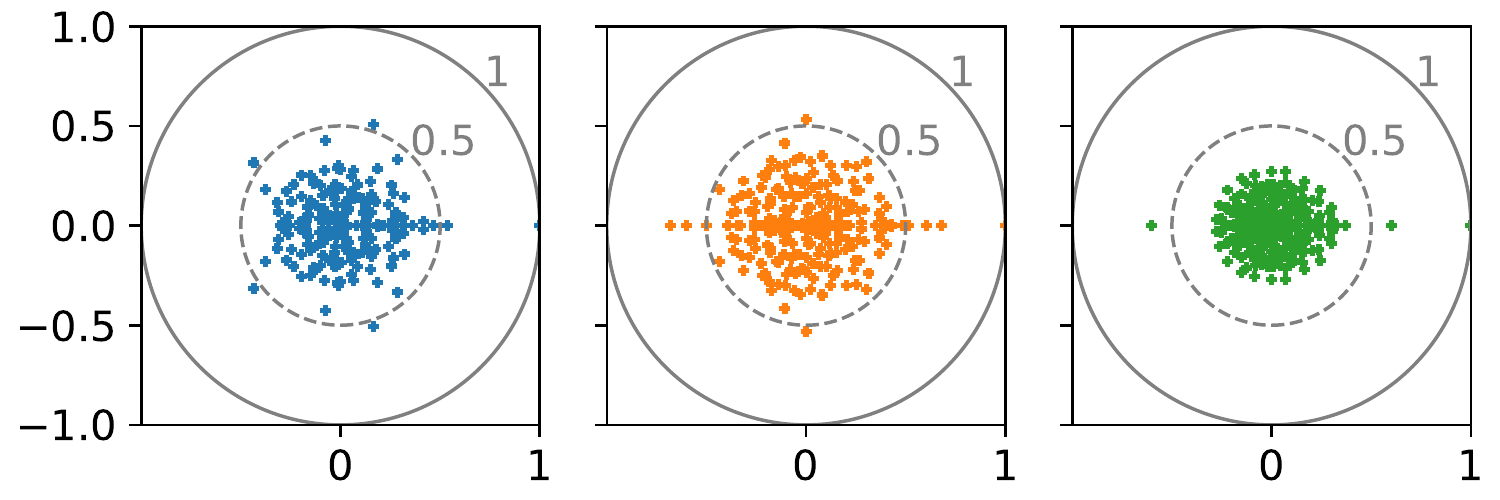}
\end{center}
\caption{From left to right: spectrum of $G$ in the complex plane for the early, middle and late periods for the major dataset.}
\label{spectreparperiode}
\end{figure}

\begin{figure*}
\begin{center}
\includegraphics[width=0.31\linewidth]{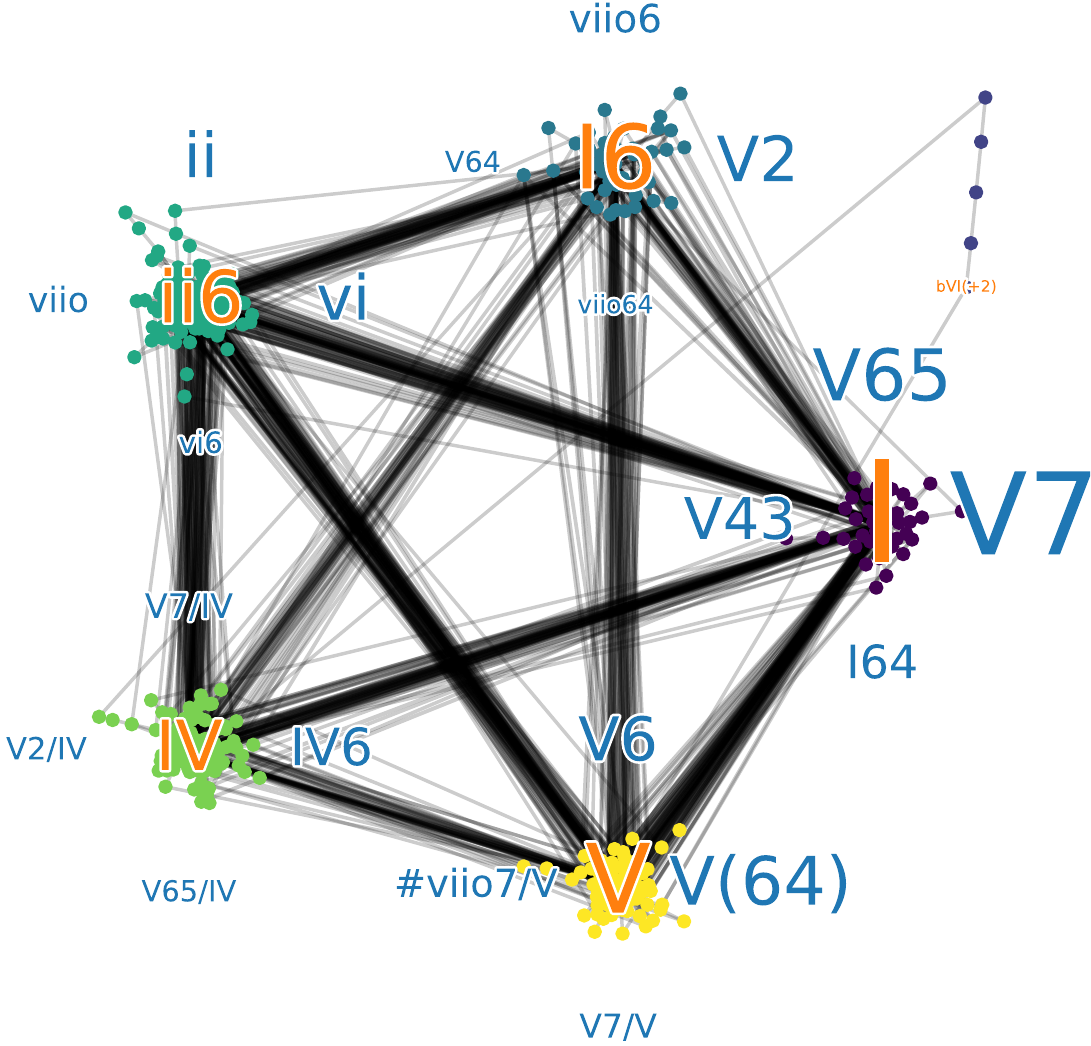}
\includegraphics[width=0.31\linewidth]{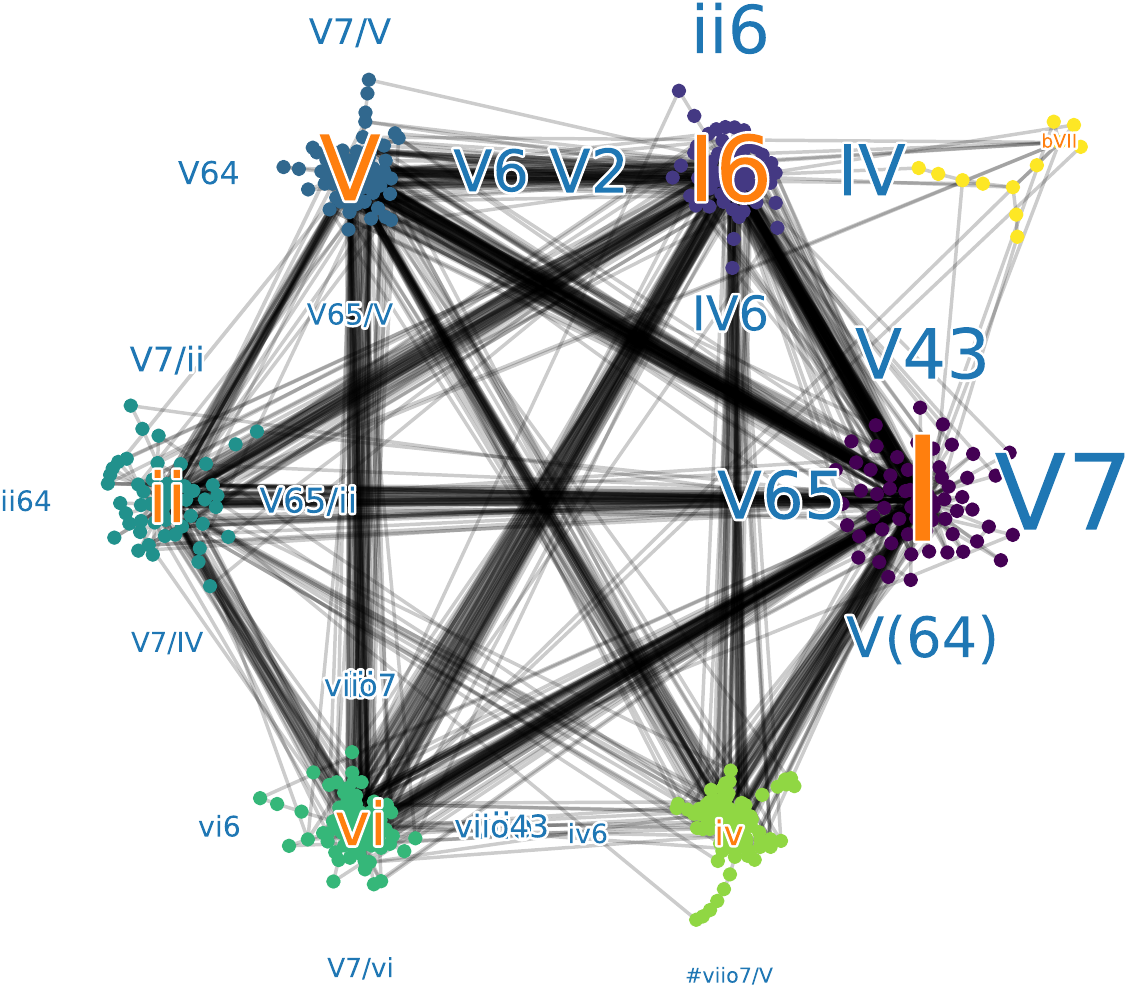}
\includegraphics[width=0.31\linewidth]{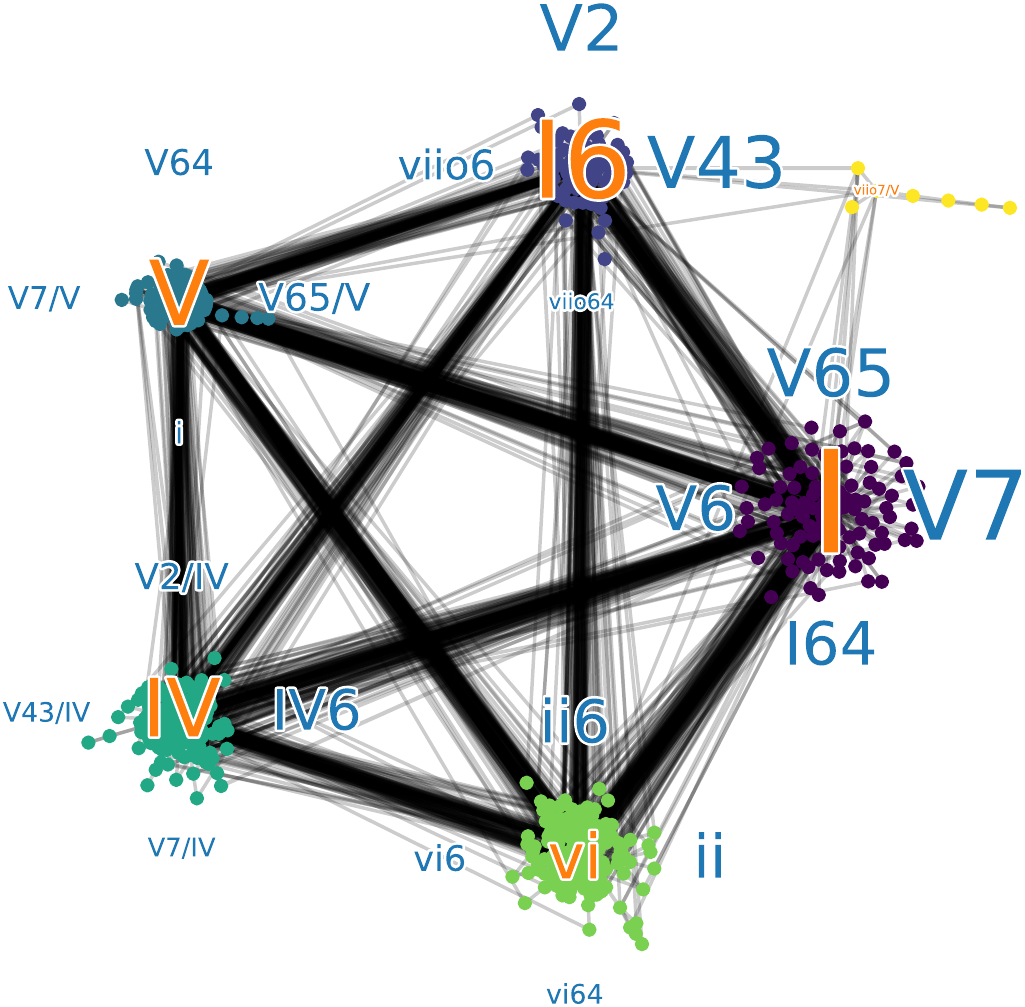}
\end{center}
\caption{Left to right: community structure of early, middle and late graphs of the major dataset, same conventions as in Fig.~\ref{figcommunautes}.}
\label{periodcommunities}
\end{figure*}

We now analyze how network properties depend on the period of composition of the quartets. It is well-known that Beethoven underwent a strong stylistic evolution during his lifetime, the last quartets in particular being more akin to romantic music that the previous ones. 
It is usually recognized that three temporally well-separated periods can be identified. The early period corresponds to the first six quartets (1798-1800), the middle period to the five next (1806-1814), and the late quartets are the last five (1824-1826), the Gro\ss{e} Fuge being the last movement of Op.\,130. In order to assess how this evolution reflects in the musical networks, for each period (early/middle/late) and each mode (major/minor) we constructed a network by only taking into account the corresponding musical segments. For major segments, the number of chords is 6496 for the early period, 4794 for the middle period, and 8986 for the late period. This results in graphs of size 379, 442 and 574, corresponding to the respective number of distinct chords.

\subsection{Spectrum}
In Fig.~\ref{spectreparperiode} we display the spectra of $G$ associated with each period for major segments.
As in Fig.~\ref{spectre} for the global network, the spectrum is concentrated near the origin, all the more so for the last period. As this concentration is linked with a strong connectivity between different parts of the network, this latter feature can be associated with the fact that the last period of Beethoven's quartets is associated with more musical originalities and formal innovation, leading to the appearance of new edges between previously unconnected chords. 
Some isolated eigenvalues can be found outside the region where most eigenvalues concentrate. In particular one can observe the presence of families of eigenvalues located at the same radius. For instance, near the dashed circle in Fig.~\ref{spectreparperiode} left are 5 eigenvalues separated by $2\pi/5$; these eigenvalues are associated with closely related eigenvectors, which are essentially located on the same chords (here, by order of amplitude, bIII, V65/bIII, iv, V6/bIII).
They correspond to groups of chords usually played together, and are relatively well separated from the rest of the network. These groups play the role of small communities. This type of eigenstates is clearly less prominent in the third network, indicating again that the last period of Beethoven's quartets contains more musical innovations, connecting isolated groups of chords in the network.

\subsection{Communities}\label{commEpoques}

The partition of the network into communities also presents different characteristics from one period to another. The community partition yielding the highest modularity over different trials of the Louvain algorithm corresponds to modularities 0.2651, 0.2607, and 0.2290, respectively. 
This is to be compared with the (maximal) modularity for graphs with same vertex degree distribution but random edges: for the corresponding sizes, modularity is respectively $0.091\pm 0.0034$ for $N=379$, $0.126\pm 0.0036$ for $N=442$, and $0.094\pm 0.0028$ for $N=574$ (the standard deviation is obtained from 1000 realizations). The decrease of modularity from early to late is therefore significant. As already pointed out above for the spectrum, this result is in line with the greater homogeneity of the late graph, due to the presence of new edges between chords not previously connected.

The community partition for the three graphs in major is displayed in Fig.~\ref{periodcommunities}. The weaker modularity of the late period is a manifestation of the larger connectivity between communities.

\subsection{Statistical analysis of the evolution of Beethoven's quartets}
\begin{figure}[!t]
\begin{center}
\includegraphics[width=1\linewidth]{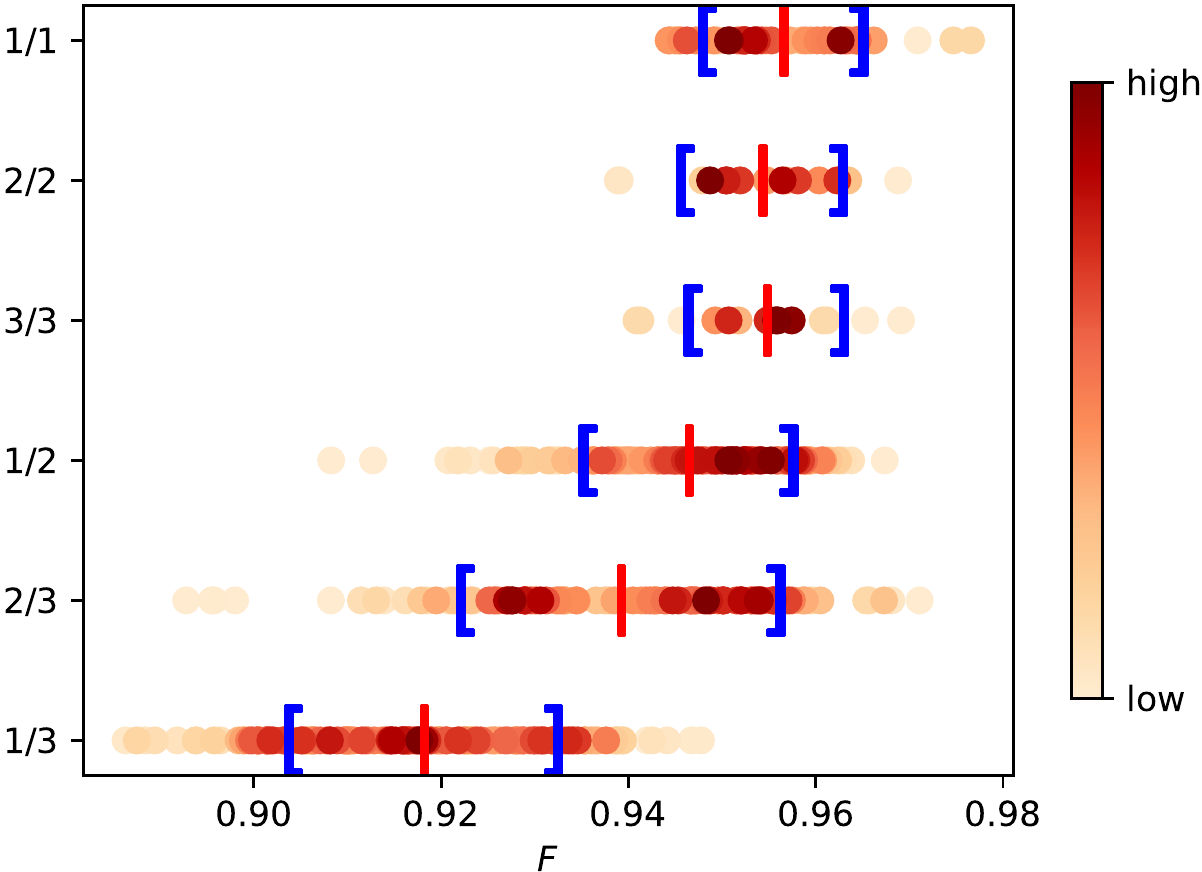}\\
\vspace{.4cm}
\includegraphics[width=1\linewidth]{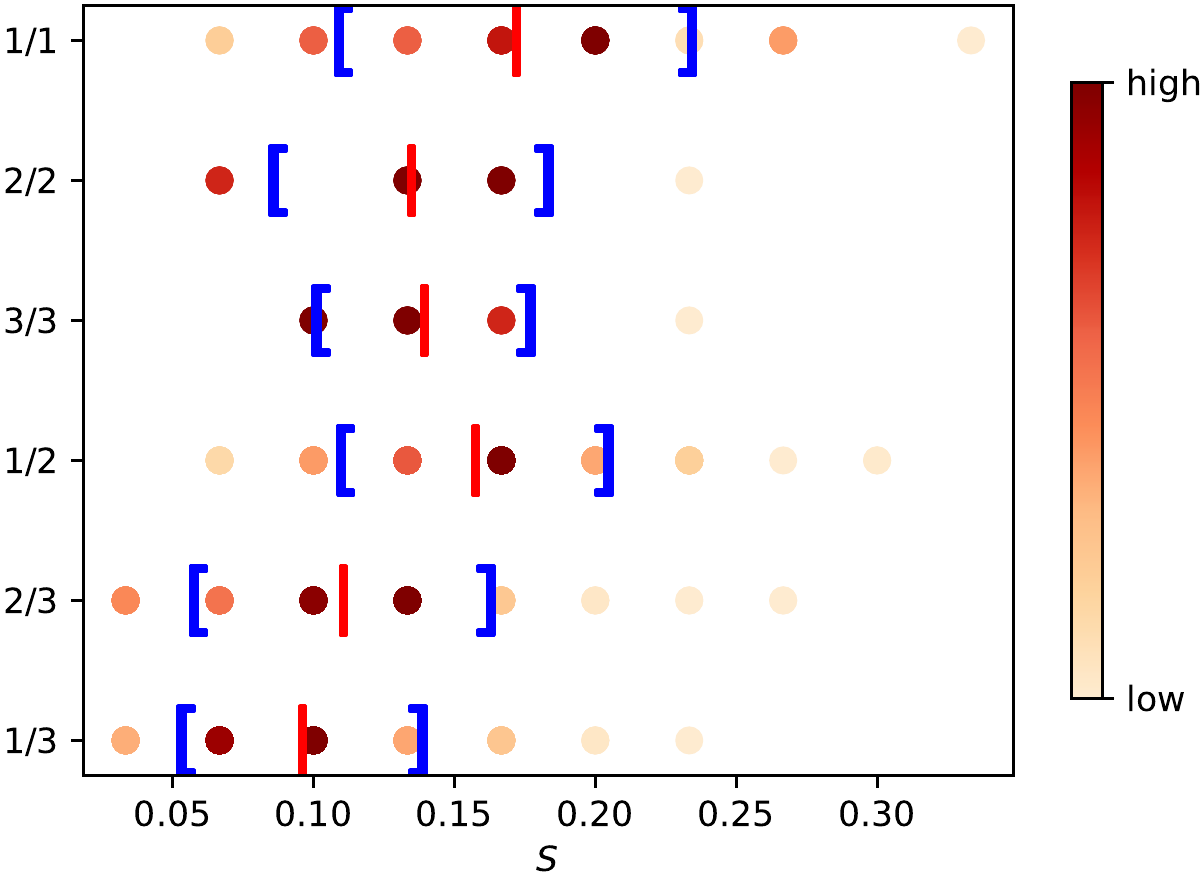}
\end{center}
\caption{Top: distribution of the fidelity $F$ for different periods.
Each point at line $i/j$ corresponds to the fidelity between a pair of distinct PageRank vectors from periods $i$ and $j$ (see text for detail). The color represents the density of points, calculated by counting the number of values within a distance of 0.002 from the actual point. Bottom: same for similarity $S$ (note that the discreteness of $S$ makes the values highly degenerate).}
\label{fidelity}
\end{figure}

We now assess whether the stylistic evolution of the quartets also reflects in statistically significant differences in the networks, using the PageRank vectors.
To analyze how close two rankings are, one can calculate the closeness of their PageRank entries. Stemming from quantum theory, the fidelity $F(\psi,\phi)$ measures how close two quantum states $\psi$ and $\phi$ are \cite{nielsen00}. It is defined as the square of the scalar product between the two vectors (normalized as  $\sum_i |\psi_i|^2=1$), namely $F(\psi,\phi)=|\sum_i\psi^*_i\phi_i|^2$. In Fig.~\ref{fidelity} we display fidelities between different PageRanks. In order to assess their statistical significance, we construct several networks for each period: namely, for any pair of quartets $a,b$ of a given period and mode, we construct the graph $\mathcal{G}_{ab}$ based on chords from quartets $a$ and $b$ only, and the corresponding PageRank vector $p$. For early major, which has 6 quartets, we get 15 different PageRank vectors, while middle and late major (with 5 quartets) have 10 PageRank vectors each. The points in Fig.~\ref{fidelity} top correspond to all possible fidelities between these vectors, with the restriction that pairs of vectors from the same period should not contain any quartet in common (for instance there are 45 such pairs when comparing the early period with itself, and 150 when comparing the early period with the middle one). 
The vertical line and parses $[\cdot|\cdot]$ indicate the mean and standard deviation of the values. 
As appears in the plot, the mean values of fidelities within a period (first three lines) are centered around 0.955. By contrast, the fidelities of pageranks from different periods are statistically weaker. The largest difference is between periods 1 and 3.

In order to compare more precisely the rankings between the different periods, we introduce the 'similarity' $S$ between vectors, which we define as the mean number of identical indices within the $m=30$ first indices when they are ordered by decreasing $p_i$. The results are plotted at Fig.~\ref{fidelity} bottom. Although the results are less statistically significant than for the fidelity, there is a clear difference between the late period and the first two, and the first and second period are much more similar. 

Our results are thus compatible with the opinion of musical scholars, showing a marked difference between the late period and the early and middle ones. This indicates that such tools from network theory are able to capture stylistic differences in music.

\section{Conclusion}

The present work shows that the complex network approach can be fruitfully applied to the harmonic structure of musical works. Based on the example of Beethoven string quartets, we specified the properties of this new type of networks, and in particular we discussed the relationship between the spectrum of the Google matrix, the community structures, and musical specificities of the scores. We have also shown that the tools of complex networks allow to distinguish between the different periods and styles of Beethoven string quartets.

Our work opens the way to similar statistical analyses for different composers. In 1815 the Allgemeine musikalische Zeitung wrote that "Beethoven is without question the boldest sailor on the tide of harmony" (cited by \cite{ratner1970key}); similar harmonic analyses of pieces by Palestrina \cite{hedges2011exploring}, Bach \cite{rohrmeier2008statistical}, Mozart \cite{hentschel2021annotated} or Schubert \cite{weiss2021schubert} from a network approach would thus give an interesting perspective. Other aspects of music, such as history of harmonic patterns \cite{anzuoni2021historical}, or rhythm \cite{rohrmeier2020towards}, could benefit from the network approach. Another possible fruitful direction  could be to apply this approach to uncover hierarchical structures in music, in the spirit of Schenkerian analysis \cite{pankhurst2008schenkerguide}, by performing some coarse-graining to the network.

\acknowledgments We thank A.~Ouali, who was involved in a preliminary study. We thank Calcul en Midi-Pyr\'en\'ees (CalMiP)  for  access  to  its  supercomputers.

\section{Appendix - Cleaning the database}
We found several issues in the database, such as missing data or faulty labels, most of which are listed at the database website \cite{abc}. We made the following corrections to the file all$\_$annotations.tsv available at \cite{abc}. Global keys labeled 'nothing' or 'false' have been replaced by their correct value, given in the first column of the database. For some entries, the local key was labeled 'Ab' instead of 'VI': they were restored to their correct value. The local keys labeled 'I' at the beginning of some minor segments were relabeled 'i'.

We also chose to consider chords within a pedal segment to be treated without reference to the pedal (although the first chord of a pedal segment is treated as distinct). As for entries labeled 'none', i.e.~chords for which no consensual harmonic analysis could be extracted from the score, we chose to treat them as a chord on its own. 

In order to  check the consistency of the corrections we made to the database, we compared our results with the ones obtained in \cite{moss2019statistical}. In particular, for both major and minor segments, we calculated the list of frequencies of each chord type and the heatmaps (frequency of each sequence of pairs of chords), following \cite{moss2019statistical}; the numerical outcomes we obtain is close to the ones obtained in \cite{moss2019statistical}. The main difference is the frequency of 'I' in minor segments, which ranks 14 in frequency order in our database but 2 in Ref.~\cite{moss2019statistical}. It is very likely that the corrections listed in \cite{abc} have been performed after \cite{moss2019statistical} was published, which would explain this discrepancy.

\bibliographystyle{eplbib}

\begin{thebibliography}{99}
\expandafter\ifx\csname url\endcsname\relax\def\url#1{\texttt{#1}}\fi

\bibitem{boccaletti2006complex}
\Name{Boccaletti S., Latora V., Moreno Y., Chavez M. \and Hwang D.-U.}
  \REVIEW{Physics Reports}{424}{2006}{175}.

\bibitem{cancho2001small}
\Name{Cancho R. F.~I. \and Sol{\'e} R.~V.} \REVIEW{Proceedings of the Royal
  Society of London. Series B: Biological Sciences}{268}{2001}{2261}.

\bibitem{masucci2006network}
\Name{Masucci A.~P. \and Rodgers G.~J.} \REVIEW{Physical Review
  E}{74}{2006}{026102}.

\bibitem{antiqueira2007strong}
\Name{Antiqueira L., Nunes M. d. G.~V., Oliveira~Jr O. \and Costa L. d.~F.}
  \REVIEW{Physica A: Statistical Mechanics and its
  Applications}{373}{2007}{811}.

\bibitem{sole2010language}
\Name{Sol{\'e} R.~V., Corominas-Murtra B., Valverde S. \and Steels L.}
  \REVIEW{Complexity}{15}{2010}{20}.

\bibitem{georgeot2012game}
\Name{Georgeot B. \and Giraud O.} \REVIEW{Europhysics
  Letters}{97}{2012}{68002}.

\bibitem{kandiah2014move}
\Name{Kandiah V., Georgeot B. \and Giraud O.} \REVIEW{The European Physical
  Journal B}{87}{2014}{1}.

\bibitem{xu2015weiqi}
\Name{Xu L.-G., Li M.-X. \and Zhou W.-X.} \REVIEW{Europhysics
  Letters}{110}{2015}{58004}.

\bibitem{coquide2017distinguishing}
\Name{Coquid{\'e} C., Georgeot B. \and Giraud O.} \REVIEW{Europhysics
  Letters}{119}{2017}{48001}.

\bibitem{benson2006music}
\Name{Benson D.} \Book{Music: A mathematical offering} (Cambridge University
  Press) 2006.

\bibitem{de2013automatic}
\Name{De~Haas W.~B., Magalh{\~a}es J.~P., Wiering F. \and Veltkamp R.~C.}
  \REVIEW{Computer Music Journal}{37}{2013}{37}.

\bibitem{kroger2008rameau}
\Name{Kr{\"o}ger P., Passos A., Sampaio M. \and De~Cidra G.} presented at
  \Book{ICMC} 2008.

\bibitem{rohrmeier2008statistical}
\Name{Rohrmeier M. \and Cross I.} in proc. of \Book{10th international
  conference on music perception and cognition} Vol.~6 (Hokkaido University
  Sapporo, Japan) 2008 pp. 619--627.

\bibitem{neuwirth2018annotated}
\Name{Neuwirth M., Harasim D., Moss F.~C. \and Rohrmeier M.} \REVIEW{Frontiers
  in Digital Humanities}{5}{2018}{16}.

\bibitem{hentschel2021annotated}
\Name{Hentschel J., Neuwirth M. \and Rohrmeier M.} \REVIEW{Transactions of the
  International Society for Music Information Retrieval}{4}{2021}{}.

\bibitem{marsden2010schenkerian}
\Name{Marsden A.} \REVIEW{Journal of New Music Research}{39}{2010}{269–289}.

\bibitem{elgammal}
\Name{Elgammal A.} \Book{Music scholars and computer scientists completed
  {B}eethoven's tenth symphony aided by machine learning}
  \url{https://tinyurl.com/f7e2mkea}.

\bibitem{landsnes2019model}
\Name{Landsnes K., Mehrabyan L., Wiklund V., Moss F.~C., Lieck R. \and
  Rohrmeier M.} in proc. of \Book{16th Sound \& Music Computing Conference}
  2019 pp. 250--254.

\bibitem{beach1974origins}
\Name{Beach D.~W.} \REVIEW{Journal of Music Theory}{18}{1974}{274}.

\bibitem{darrigol2007acoustic}
\Name{Darrigol O.} \REVIEW{Archive for history of exact
  sciences}{61}{2007}{343}.

\bibitem{moss2019statistical}
\Name{Moss F.~C., Neuwirth M., Harasim D. \and Rohrmeier M.} \REVIEW{PloS
  one}{14}{2019}{e0217242}.

\bibitem{ratner1970key}
\Name{Ratner L.~G.} \REVIEW{J.~Am.~Music.~Soc.}{23}{1970}{472}.

\bibitem{knittel1998wagner}
\Name{Knittel K.~M.} \REVIEW{J.~Am.~Music.~Soc.}{51}{1998}{49}.

\bibitem{bonds2017irony}
\Name{Bonds M.~E.} \REVIEW{J.~Am.~Music.~Soc.}{70}{2017}{285}.

\bibitem{tymoczko2006geometry}
\Name{Tymoczko D.} \REVIEW{Science}{313}{2006}{72}.

\bibitem{buongiorno2020hitchhiker}
\Name{Buongiorno~Nardelli M.} \REVIEW{arXiv e-print 2006.05007}{}{2020}{}.

\bibitem{buongiorno2021tonal}
\Name{Buongiorno~Nardelli M.} \REVIEW{Journal of Mathematics and
  Music}{}{2021}{1}.

\bibitem{abc}
\Name{Neuwirth M., Harasim D., Moss F.~C. \and Rohrmeier M.}
  \url{https://github.com/DCMLab/ABC}.

\bibitem{zipf1935psycho}
\Name{Zipf G.} \Book{The psycho-biology of language: an introduction to dynamic
  philology.} (1935).

\bibitem{musescore}
\Book{Muse{S}core} version 3.6.2, released under the GNU GPLv2 license.

\bibitem{brin1998anatomy}
\Name{Brin S. \and Page L.} \REVIEW{Computer networks and ISDN
  systems}{30}{1998}{107}.

\bibitem{ermann2015google}
\Name{Ermann L., Frahm K.~M. \and Shepelyansky D.~L.} \REVIEW{Reviews of Modern
  Physics}{87}{2015}{1261}.

\bibitem{newman2004finding}
\Name{Newman M.~E. \and Girvan M.} \REVIEW{Physical Review
  E}{69}{2004}{026113}.

\bibitem{fortunato2010community}
\Name{Fortunato S.} \REVIEW{Physics Reports}{486}{2010}{75}.

\bibitem{barabasi1999emergence}
\Name{Barab{\'a}si A.-L. \and Albert R.} \REVIEW{Science}{286}{1999}{509}.

\bibitem{albert2002statistical}
\Name{Albert R. \and Barab{\'a}si A.-L.} \REVIEW{Review of Modern
  Physics}{74}{2002}{47}.

\bibitem{ebel2002scale}
\Name{Ebel H., Mielsch L.-I. \and Bornholdt S.} \REVIEW{Physical Review
  E}{66}{2002}{035103}.

\bibitem{donato2004large}
\Name{Donato D., Laura L., Leonardi S. \and Millozzi S.} \REVIEW{The European
  Physical Journal B}{38}{2004}{239}.

\bibitem{pandurangan2006using}
\Name{Pandurangan G., Raghavan P. \and Upfal E.} \REVIEW{Internet
  Mathematics}{3}{2006}{1}.

\bibitem{giraud2009delocalization}
\Name{Giraud O., Georgeot B. \and Shepelyansky D.~L.} \REVIEW{Physical Review
  E}{80}{2009}{026107}.

\bibitem{georgeot2010spectral}
\Name{Georgeot B., Giraud O. \and Shepelyansky D.~L.} \REVIEW{Physical Review
  E}{81}{2010}{056109}.

\bibitem{blondel2008fast}
\Name{Blondel V.~D., Guillaume J.-L., Lambiotte R. \and Lefebvre E.}
  \REVIEW{Journal of Statistical Mechanics: theory and
  experiment}{2008}{2008}{P10008}.

\bibitem{nielsen00}
\Name{Nielsen M.~A. \and Chuang I.~L.} \Book{Quantum Computation and Quantum
  Information} (Cambridge University Press) 2000.

\bibitem{hedges2011exploring}
\Name{Hedges T. \and Rohrmeier M.} in proc. of \Book{International Conference
  on Mathematics and Computation in Music} (Springer) 2011 pp. 334--337.

\bibitem{weiss2021schubert}
\Name{Wei{\ss} C., Zalkow F., Arifi-M{\"u}ller V., M{\"u}ller M., Koops H.~V.,
  Volk A. \and Grohganz H.~G.} \REVIEW{Journal on Computing and Cultural
  Heritage}{14}{2021}{1}.

\bibitem{anzuoni2021historical}
\Name{Anzuoni E., Ayhan S., Dutto F., McLeod A., Moss F.~C. \and Rohrmeier M.}
  in proc. of \Book{18th Sound and Music Computing Conference} 2021 pp.
  284--291.

\bibitem{rohrmeier2020towards}
\Name{Rohrmeier M.} presented at \Book{21st Int. Society for Music Information
  Retrieval Conference} 2020.

\bibitem{pankhurst2008schenkerguide}
\Name{Pankhurst T.} \Book{SchenkerGUIDE: a brief handbook and website for
  Schenkerian analysis} (Routledge) 2008.

\end{thebibliography}

\end{document}